\documentclass[aps,prb,twocolumn,groupedaddress,showpacs]{revtex4}
\usepackage{bm}
\begin{document}
\title{Coupling of Josephson current qubits using a connecting loop}
\author{Mun Dae Kim \cite{Corresp} and Jongbae Hong}
\affiliation{School of Physics, Seoul National University, Seoul 151-742, Korea}
\begin{abstract}
We propose a coupling scheme for the three-Josephson junction qubits
which uses a connecting loop, but not mutual inductance.
Present scheme offers the advantages of a large and tunable level splitting
in implementing the controlled-NOT (CNOT) operation. We calculate the switching
probabilities of the coupled qubits in the CNOT operations
and demonstrate that present CNOT gate can meet the criteria for
the fault-tolerant quantum computing.
We obtain the coupling strength as a function of
the  coupling energy of the Josephson junction and the length of the connecting loop
which varies with selecting two qubits from  the scalable design.
\end{abstract}
\draft
\maketitle

\newpage

\section{Introduction}

The persistent current qubit using superconducting loop with three Josephson junctions
has been proposed as a promising candidate for quantum computer
\cite{Mooij,Orlando,Wal,Chio} owing to
the advantages of relatively long decoherence time and possible scalability.
The current qubit has shown quantum superposition\cite{Wal}
and, recently,  coherent time evolution\cite{Chio}
between two quantum states. Therefore, the feasibility of
the current qubit as a practical quantum computer has increased recently.
In order to realize a practical quantum computer,
qubits must be coupled with scalability and
any selected pair of qubits should perform a universal
logic gate. One of universal logic gates is
composed of the controlled-NOT (CNOT) operations and the single-qubit
rotations.\cite{Nielsen,Galindo}

In this study we suggest a coupling scheme for  CNOT
operation  and  scalable design using the
three-junction current qubit  as our unit qubit whose feasibility
as a single qubit has already been tested.\cite{Mooij,Orlando,Wal,Chio}
Here we propose a design for coupling two qubits using a connecting loop.
Two qubits are connected by a superconducting loop with Josephson junctions which
carries a persistent current by  effective piercing flux.
The effective flux comes from the phase differences across the Josephson junctions
common to the qubit loop and the connecting loop.
We calculate coupling strength between two qubits as a function of
the coupling energy of Josephson junction in connecting loop and
the length of connecting loop which varies with  selecting qubits
from the scalable circuit.

Recent experiments on two coupled charge qubits have demonstrated
coherent oscillation\cite{Pashkin} and CNOT gate operation.\cite{Yamamoto}
For the coupled three-Josephson junction qubits\cite{Majer} and the coupled current-biased
Josephson junction qubits\cite{Berkley} spectroscopy measurements have been performed,
but the CNOT operation has not yet been demonstrated.
It seems that, for the coupled three-Josephson junction qubits, the coupling strength
between qubits is too weak to implement CNOT operation.
Here we aim to propose  a coupling scheme which is expected to have strong
advantages in CNOT operation, since the coupling
strength can be large enough to suppress the unwanted oscillation of the control qubit.
In order to obtain  quantitative results we calculate the switching probabilities
of the coupled qubit system in CNOT operation. For an appropriate range of parameters,
target qubit oscillates and shifts to another state while the unwanted oscillation
of control qubit can be suppressed, which can be implemented within the fault-tolerant
criteria of quantum computing.\cite{Nielsen,Galindo}
Moreover present coupling scheme offers the tunable coupling strength and the
selective coupling from the scalable design, which is crucial for implementation
of practical quantum computing but is not provided by the previous coupling schemes.
\cite{Pashkin,Yamamoto,Majer,Berkley}
We also show that the coupling strength does not diminish rapidly
as length of the connecting loop increases.

\section{Hamiltonian of the coupled qubits}

\begin{figure}[b]
\vspace*{6cm}
\includegraphics{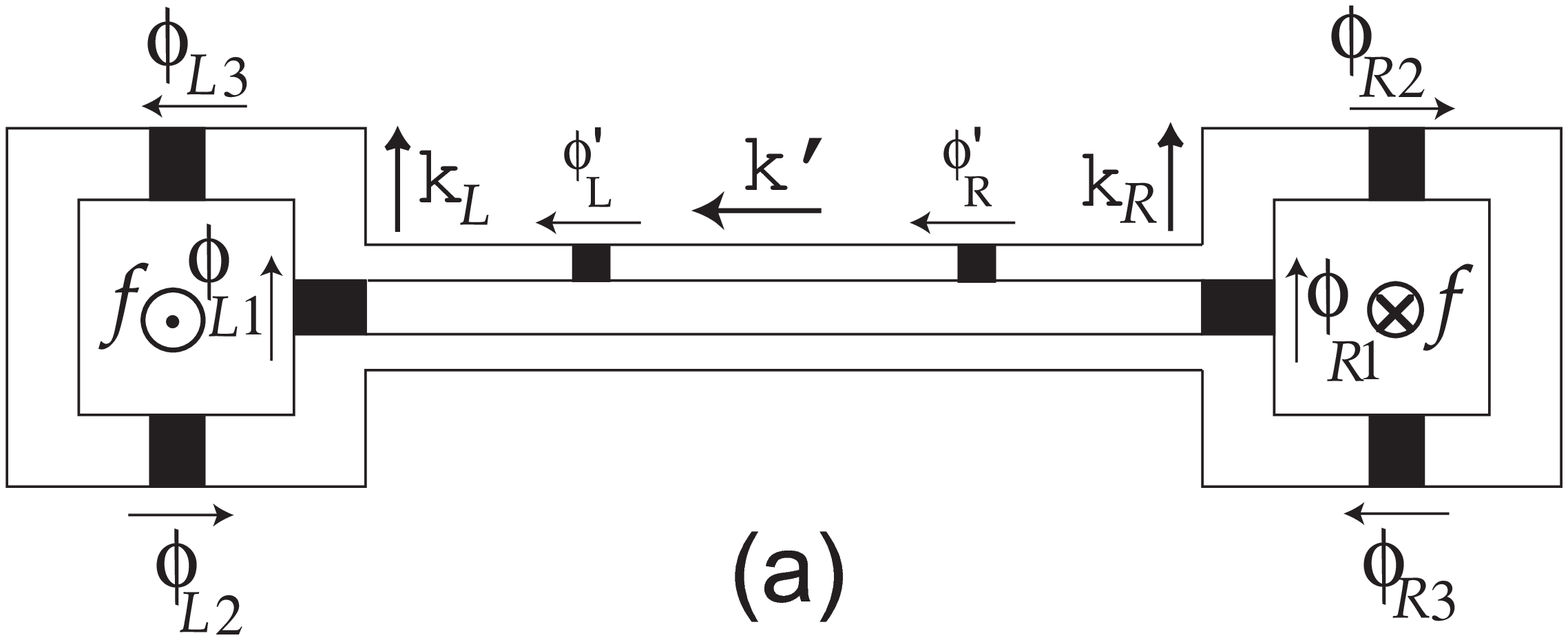}
\vspace*{0cm}
\includegraphics{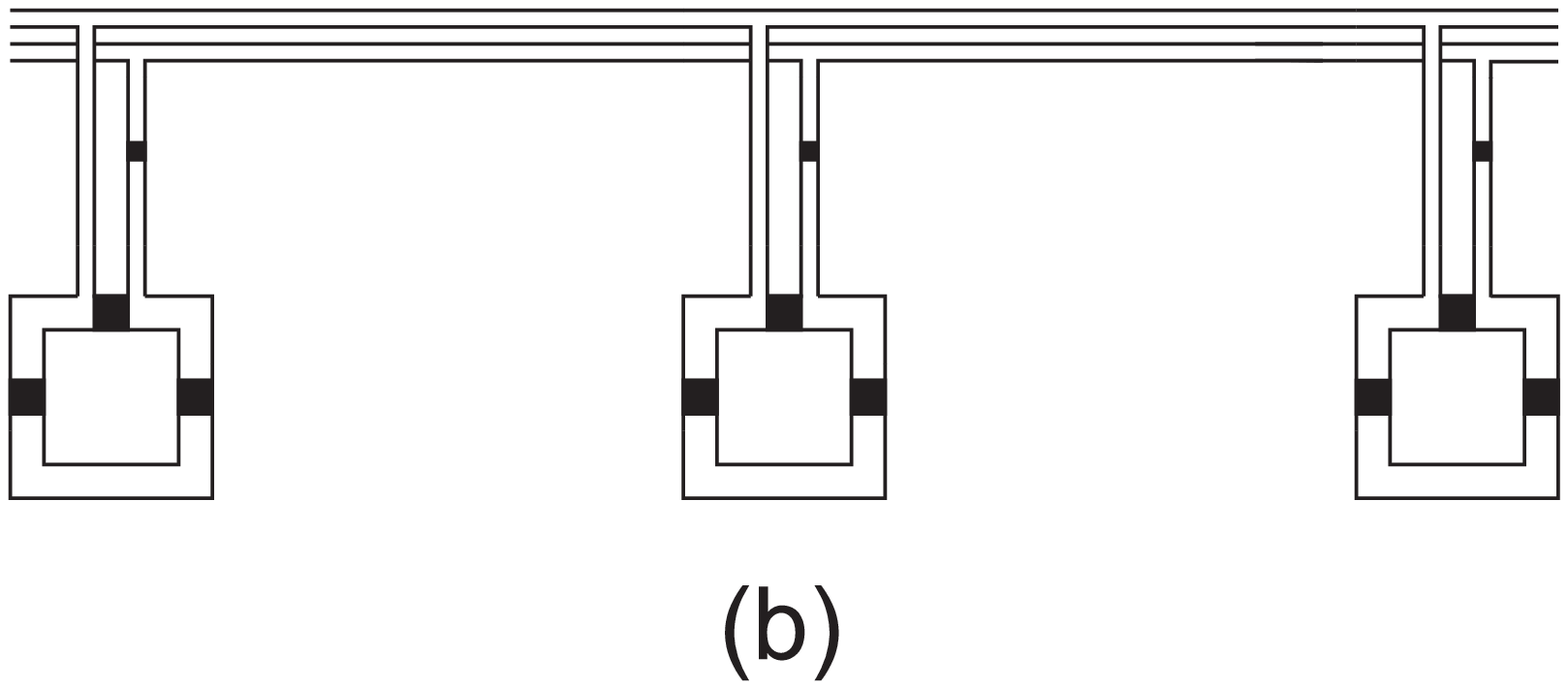}
\vspace*{1cm}
\caption{(a) A  pair of coupled qubits.
(b) A part of the scalable quantum computing device. One unit is
composed of a three-junction qubit, a dc-SQUID, and a connecting loop.}
\label{fig:fig1}
\end{figure}

A  pair of coupled qubits  is drawn in Fig. 1 (a). Two qubits are coupled through a connecting loop.
Using this coupled qubits as a basic unit, a possible design to achieve a scalable computing
is presented in Fig. 1 (b). We express the direction of the penetrating flux of the qubits oppositely
in Fig. 1 (a), because two selected qubits are connected in a twisted way as shown in Fig. 1 (b).
The two Josephson junctions in the connecting loop of Fig. 1 (a) denote the dc-SQUID's
that enable tunable coupling and switching function for selective coupling.
The wiring scheme connecting any pair of qubits is topologically the same.
The interqubit magnetic crosstalk between the qubit islands
can be neglected, since the induced flux of the three-Josephson
junction loop is very weak and the qubit islands are separated far enough.
In Fig. 1 (a) the weak magnetic interaction between the qubit loop and the connecting loop
may exist even though the area of the connecting loop is  very small.
However, by manufacturing the connecting loop in
such a way that the normal directions of the qubit loop and the connecting loop are
perpendicular to each other, we can  prevent this unwanted interaction.

The fluxoid quantization  condition for  a thin superconducting loop
\cite{Tinkham},
\begin{equation}
\label{fluxoid}
-\Phi_t+(m_c/q_c)\oint \vec{v}_c \cdot d\vec{l}=n\Phi_0
\end{equation}
with $q_c=2e$, $m_c=2m_e$, $\vec{v}_c$  the average velocity of Cooper pairs,
$\Phi_0\equiv h/2e$ the superconducting unit flux quantum and $\Phi_t$ the total flux,
can be represented as the  periodic boundary condition for the wave function of the
Cooper pairs including the phase evolution accumulated
over the circumference of the loop with three Josephson junctions
as follows,\cite{Kim,Matveev}
\begin{equation}
\label{kn} k_nl=2\pi n+2\pi f_t-(\phi_1+\phi_2+\phi_3),
\end{equation}
where  $k_n$ is the wave vector of the Cooper pairs, $n$ an integer, $l$  the
circumference of the loop, and $f_t\equiv\Phi_{t}/\Phi_0$.
Here $\phi_i$ is the phase difference across the $i$-th Josephson junction in a qubit.

Since the total flux is the sum of external and induced one, it is
written as
\begin{eqnarray}
\label{ft} f_t=f+\frac{L_sI}{\Phi_0}=f-\frac{\gamma l}{2\pi}k_n,
\end{eqnarray}
where  $f\equiv \Phi_{\rm ext}/\Phi_0$ and
$\gamma\equiv L_s/L_K$.  Here $L_s$ is the self inductance  and
$L_K \equiv m_cl/An_cq_c^2$ the kinetic inductance of the loop,
where $n_c$ is  the density of Cooper pairs, $A$ the  cross
section of the ring, and $m_c=2m_e$ the  mass of a Cooper pair.
The dimensionless parameter $\gamma$ can also be represented by the ratio
of the energy scale such as $\gamma=E_K/E_L$ with $E_K \equiv \Phi^2_0/2L_K$
and $E_L \equiv \Phi^2_0/2L_s$.
Here use of the current expression $I=-n_cAq_c\hbar k_n/m_c$ has been made.
Then $k_nl$ of Eq. (\ref{kn}) is reexpressed as
\begin{eqnarray}
\label{kn2}
(1+\gamma)k_{R(L)}l= ~~~~~~~~~~~~~~~~~~~~~~~~~~~~~~~~~~~~~~~~~~~~~~~\nonumber\\
2\pi n_{R(L)}+2\pi f_{R(L)}-(\phi_{R(L)1}+\phi_{R(L)2}+\phi_{R(L)3}),
\end{eqnarray}
where $R(L)$ denotes the right (left) qubit and $k_{R(L)}$ is $k_nl$ of the right (left) qubit.

Since the value of $\gamma$ for the three-Josephson junction qubit loop \cite{Wal}
is very large such that $\gamma\approx 200$, Eq. (\ref{kn2}) can be written as
$\gamma k_{R(L)}l \approx 2\pi n_{R(L)}+2\pi f_{R(L)}-(\phi_{R(L)1}+\phi_{R(L)2}+\phi_{R(L)3})$
and the boundary condition in Eq. (\ref{kn}) becomes
\begin{eqnarray}
2\pi n+2\pi f_t-(\phi_1+\phi_2+\phi_3)\approx 0.
\end{eqnarray}
Using the relation in Eq. (\ref{ft}) we can obtain the relation for the
current in the qubit loop given by
\begin{eqnarray}
\label{qubitC}
I_{R(L)}=~~~~~~~~~~~~~~~~~~~~~~~~~~~~~~~~~~~~~~~~~~~~~~~~~~~~~~~~~~\nonumber\\
-\frac{\Phi_0}{L_s}
\left(n_{R(L)}+f_{R(L)}-\frac{\phi_{R(L)1}+\phi_{R(L)2}+\phi_{R(L)3}}{2\pi}\right).
\end{eqnarray}
From this expression we can get for the effective potential $U_{\rm eff} (\phi)$,
\begin{eqnarray}
U^{R(L)}_{\rm eff} (\phi_{R(L)1},\phi_{R(L)2},\phi_{R(L)3})~~~~~~~~~~~~~~~~~~~~~~~~\nonumber\\
=\sum^3_{i=1} E_{Ji}(1-\cos\phi_{R(L)i})+E^{R(L)}_n,~~~~~~
\end{eqnarray}
where $E_{Ji}=E_{JRi}=E_{JLi}$ is the Josephson coupling energy and
$E^{R(L)}_n=I^2_{R(L)}/2L_s$ can be represented  as
\begin{eqnarray}
E^{R(L)}_n=~~~~~~~~~~~~~~~~~~~~~~~~~~~~~~~~~~~~~~~~~~~~~~~~~~~~~~~~~~\nonumber\\
\frac{\Phi^2_0}{2L_s}\left[n_{R(L)}+f_{R(L)}-\frac{\phi_{R(L)1}+\phi_{R(L)2}+\phi_{R(L)3}}{2\pi}\right]^2.
\end{eqnarray}

In order to obtain the effective potential of  connecting loop, we
use the  periodic boundary condition over the circumference of
connecting loop with  self inductance $L'_s$ and obtain the relation for
the current in connecting loop such that
\begin{eqnarray}
\label{loopC}
I'=-\frac{\Phi_0}{L'_s}\left(r-\frac{\phi'_{L}+\phi'_{R}+\phi_{R1}-\phi_{L1}}{2\pi}\right),
\end{eqnarray}
where $r$ is an integer.
The superscript prime in this work means the connecting loop.
The  effective potential of connecting loop with the Josephson coupling energy
$E'_J$ can be represented by
\begin{eqnarray}
U'_{\rm eff}(\phi'_L,\phi'_R,\phi_{R1},\phi_{L1})~~~~~~~~~~~~~~~~~~~~~~~~~~~~~~~~~\nonumber\\
=E'_n+E'_{J}(1-\cos\phi'_L)+E'_{J}(1-\cos\phi'_R),
\end{eqnarray}
where
\begin{eqnarray}
\label{deltaE}
E'_n=\frac{\Phi^2_0}{2L'_s}\left[r-\frac{\phi'_L+\phi'_R+\phi_{R1}-\phi_{L1}}{2\pi}\right]^2.
\end{eqnarray}
The total effective potential,
$U^{\rm tot}_{\rm eff}$, of the coupled qubit system is given by
\begin{eqnarray}
U^{\rm tot}_{\rm eff}(\bm{\phi})&=&
U^L_{\rm eff}(\phi_{L1},\phi_{L2},\phi_{L3})+U^R_{\rm eff}(\phi_{R1},\phi_{R2},\phi_{R3})
\nonumber\\
&+&U'_{\rm eff}(\phi'_L,\phi'_R,\phi_{R1},\phi_{L1}),
\end{eqnarray}
where $\bm{\phi}=(\phi_{L1}, \phi_{L2}, \phi_{L3}, \phi_{R1}, \phi_{R2}, \phi_{R3}, \phi'_L, \phi'_R)$.

The local minima of the total effective potential can be obtained
by minimizing the effective potential $U^{\rm tot}_{\rm eff}(\bm{\phi})$
with respect to $\phi_i$, i.e., $\partial U_{{\rm eff}}^{\rm tot}/\partial \phi_i=0$,
which results the current relations at the Josephson junctions as follows:
\begin{eqnarray}
\label{current1}
I'+I_R+\frac{2\pi E_{J1}}{\Phi_0}\sin\phi_{R1}&=&0,\\
\label{current2}
-I'+I_L+\frac{2\pi E_{J1}}{\Phi_0}\sin\phi_{L1}&=&0,\\
\label{current3}
I_{R(L)}+\frac{2\pi E_{Ji}}{\Phi_0}\sin\phi_{R(L)i}&=&0,\\
\label{current4}
I'+\frac{2\pi E'_J}{\Phi_0}\sin\phi'_{R(L)}&=&0,
\end{eqnarray}
where $i$=2,3.

The energy of  connecting loop depends on  current state of
each qubit, i.e., diamagnetic or paramagnetic. We
express the diamagnetic persistent current state as
$|\downarrow\rangle$ and the paramagnetic state as
$|\uparrow\rangle$ using spin language. In this scheme one of
three junctions in each qubit is included in the connecting loop
as can be seen in Fig. 1 (a). The phases of each qubit,
$\phi_{L1}$ and $\phi_{R1}$, determine the energy of coupled
qubits. Sign of these phases  is positive (negative) for
diamagnetic (paramagnetic) persistent current. The value of
$(\phi_{R1}-\phi_{L1})$ in the periodic boundary condition
of Eq. (\ref{loopC}) for the connecting loop plays the role of
effective external flux in the connecting loop. When one qubit is
in the paramagnetic current state and the other in the diamagnetic
state, i.e., $|\downarrow\uparrow\rangle$ or
$|\uparrow\downarrow\rangle$, the phases $\phi_{R1}$ and
$\phi_{L1}$ have  opposite sign and induce a large value of
$(\phi_{R1}-\phi_{L1})$. In this case, the phase difference $\phi'_L$ and
$\phi'_R$ induced by  this effective flux gives the large Josephson coupling
energy in the connecting loop. However, when both qubits
are in the same current state, i.e.,
$|\downarrow\downarrow\rangle$ or $|\uparrow\uparrow\rangle$, the
phases $\phi_{R1}$ and $\phi_{L1}$ have the same sign and are
canceled in $(\phi_{R1}-\phi_{L1})$. Then the energy  of
connecting loop become very small.

The charging energies of  Josephson junctions give the kinetic part of
Hamiltonian. The number of excess Cooper pair charges on Josephson junction
$\hat{N_i}\equiv \hat{Q}_i/q_c$ is conjugate to the phase difference $\hat{\phi}_i$
and  the commutation relation $[\hat{\phi}_i,\hat{N_i}]=i$ gives the quantum phase fluctuations
of the junction. Using the Josephson relation, $Q_i=C(\Phi_0/2\pi){\dot\phi_i}$,
charging energy, $E_C=\sum_iQ_i^2/2C_i$, becomes
\begin{eqnarray}
E_C=\frac12 \left(\frac{\Phi_0}{2\pi}\right)^2\sum_{P=L,R}
\left(\sum^3_{i=1}C_{Pi}\dot{\phi}^2_{Pi}+C'_P\dot{\phi'}_P^2 \right),
\end{eqnarray}
where $C_{L(R)i}$ and $C'_{L(R)}$ are the capacitances of the Josephson junctions of the qubit
loop and the connecting loop, respectively.
If we introduce the canonical momentum
$\hat{P}_i\equiv \hat{N_i}\hbar= -i\hbar\partial /\partial \hat{\phi}_i$
and the effective mass $M_{ij}=(\Phi_0/2\pi)^2C_i\delta_{ij}$,
the  Hamiltonian can be given by
\begin{eqnarray}
\hat{H}=\frac12\hat{{\bf P}}^T\cdot {\bf M}^{-1} \cdot \hat{{\bf P}}
+U^{\rm tot}_{\rm eff}(\hat{\bm{\phi}}),
\end{eqnarray}
which describes the dynamics of the particle with effective mass
$M$ in the effective potential $U^{\rm tot}_{\rm eff}(\hat{\bm{\phi}})$.\cite{Orlando}

For a given value of the external flux the effective potential of a three-Josephson
junction qubit has a double well structure corresponding to the diamagnetic state,
$|\downarrow\rangle$, and the paramagnetic state, $|\uparrow\rangle$ at the potential minima.
The diamagnetic state and the paramagnetic state have the energy difference
depending on the value of the external flux.
The charging energy $E_C$ causes quantum tunneling through the barrier of
the double well potential resulting the quantum mechanically superposed states.
The Hamiltonian describing a qubit then becomes
$H_{\rm qubit}=h_0\sigma^z+t_0\sigma^x$, where $2h_0$ is the energy
difference between two states, $|\downarrow\rangle$ and
$|\uparrow\rangle$, and $t_0$ is the tunneling amplitude.

\begin{figure}[b]
\vspace*{8.5cm}
\includegraphics{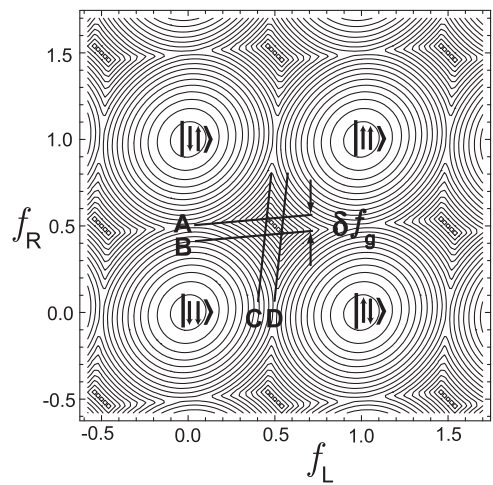}
\vspace*{-1.5cm}
\caption{ Ground state energy diagram for coupled current qubits.
A,B,C, and D denote the resonance lines and the external flux difference
between resonance lines, $\delta f_g$, originates from the coupling energy.
We consider $E'_J=E_J$.}
\label{fig:fig2}
\end{figure}

For the coupled two qubits, on the other hand, we have four states,
$|\downarrow\downarrow\rangle$, $|\downarrow\uparrow\rangle$,
$|\uparrow\downarrow\rangle$ and $|\uparrow\uparrow\rangle$,
at the effective potential minima for a given value of $(f_L,f_R)$.
Solving coupled equations (\ref{current1})$-$(\ref{current4}) in conjunction with
Eqs. (\ref{qubitC}) and (\ref{loopC}) numerically, we obtain  the local minima of
the effective potential  of the coupled qubits.
We set $\gamma = 200$, $L_s$=11pH and $E_{Ji}$=200GHz, 
which are the design parameters of the qubit of Ref. 3.
For the connecting loop we set $\gamma'=\gamma$ and $L'_s=5L_s$.
Obtaining the effective potential minima, $U^{\rm tot}_{\rm eff}(\bm{\phi}_{s,\rm min})$,
we get the energy levels at each potential well,
\begin{eqnarray}
\label{Es}
E_s=\frac12\hbar\omega_s+U^{\rm tot}_{\rm eff}(\bm{\phi}_{s,\rm min}),
\end{eqnarray}
where $s$ denotes the four states, $|\downarrow\downarrow\rangle$, $|\uparrow\downarrow\rangle$,
$|\downarrow\uparrow\rangle$  and $|\uparrow\uparrow\rangle$, and $\bm{\phi}_{s,\rm min}$ is
the value of $\bm{\phi}_{s}$ at the potential minimum corresponding to the state $s$.
The ground state energy, $(1/2)\hbar\omega_s$, can be obtained
in harmonic oscillator approximation.\cite{Orlando}
Since the plasma frequency of the Josephson junction of the three-Josephson junction loop,
$\omega_i=\sqrt{E_{Ji}/(\Phi_0/2\pi)^2 C}$, is larger than the characteristic
frequencies, i.e. the level spacing and the amplitude of the tunneling
between two states $|\uparrow\rangle$ and $|\downarrow\rangle$, by order of two \cite{Orlando}
and the plasma frequency of the Josephson junctions in the connecting loop can be the same
magnitude as that of the qubit loop, we can consider only the ground state energy.
By calculating the energy $E_s$ in Eq. (\ref{Es}) for a certain range of $(f_L,f_R)$,
we can obtain four energy planes.
In Fig. \ref{fig:fig2} we show the lowest energy  planes and
the states of the coupled qubits corresponding to the energy levels.
The finite value of $\delta f_g$ comes from the coupling energy
which is higher for different qubit state than the same qubit state.
Similar diagrams  are shown for coupled charge qubits.\cite{Pashkin,Yamamoto,Averin}

%

\begin{figure}[b]
\vspace*{6.3cm}
\includegraphics{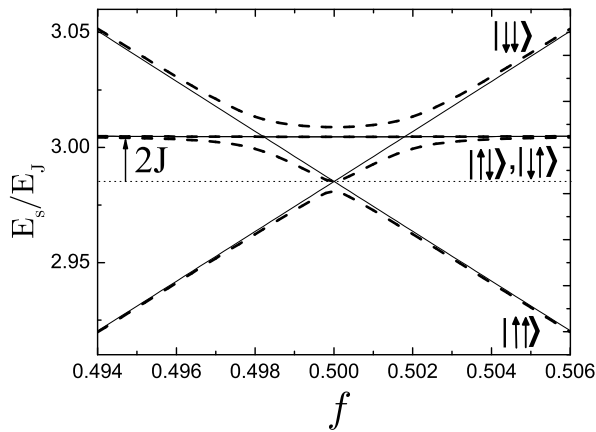}
\vspace*{-0.5cm}
\caption{Energy diagram  of the current states of coupled qubits,
when $f_L=f_R=f$ and $E'_J=0.02E_J$.
The energies of the coupled qubits without tunneling are represented by solid lines.
Dashed lines denote the eigenvalues of $H_{\rm coup}$ in Eq. (\ref{totalH}) with $t_L=t_R$=1GHz.
The energy of antiparallel spin state of coupled qubits shifts upward by amount of
$2J$ compared with that of the uncoupled state shown as dotted line.}
\label{fig:fig3}
\end{figure}

The Hamiltonian for the coupled qubits  can be written
in the two-qubit  basis, $|\downarrow\downarrow\rangle, |\uparrow\downarrow\rangle,
|\downarrow\uparrow\rangle,$ and $|\uparrow\uparrow\rangle$, \cite{Pashkin}
from the calculated values of $E_s$ in Eq. (\ref{Es}) as follows,
\begin{eqnarray}
\label{totalH}
H_{\rm coup}=\left(\matrix{E_{\downarrow\downarrow} & t_L & t_R & 0 \cr
t_L & E_{\uparrow\downarrow} & 0 & t_R \cr
t_R & 0 & E_{\downarrow\uparrow} & t_L \cr
0 & t_R & t_L & E_{\uparrow\uparrow}}\right),
\end{eqnarray}
where $t_{L(R)}$ is the tunneling amplitude across the barrier of  the double well
potential of the left (right) qubit and can be calculated by various methods.
\cite{Orlando,Kim}
These basis states we use do not directly correspond to the individual
states without coupling, but to the coupled qubit states characterized by the
current states of the two qubits and the connecting loop.
A straightforward calculation leads another form of the Hamiltonian $H_{\rm coup}$,
\begin{eqnarray}
\label{Hcoup}
H_{\rm coup}&=&h_L\sigma^z_L\otimes I +h_R I\otimes\sigma^z_R -J\sigma^z_L\otimes \sigma^z_R
\nonumber\\
&+&t_L\sigma^x_L\otimes I +t_R I\otimes\sigma^x_R
\end{eqnarray}
with relations that
$E_{\downarrow\downarrow}=h_L+h_R-J, E_{\uparrow\downarrow}=-h_L+h_R+J,
E_{\downarrow\uparrow}=h_L-h_R+J$ and $E_{\uparrow\uparrow}=-h_L-h_R-J.$
Here $I$ is the $2\times 2$ identity matrix and the coupling constant $J$ can be represented as
\begin{eqnarray}
\label{J}
J=\frac14(E_{\downarrow\uparrow}
+E_{\uparrow\downarrow}-E_{\downarrow\downarrow}-E_{\uparrow\uparrow}),
\end{eqnarray}
which shows that $J$ is the  energy difference between the parallel spin states
and the antiparallel spin states as shown in Fig. \ref{fig:fig3}.

Figure \ref{fig:fig3} shows the energy levels of Fig. \ref{fig:fig2}
along the symmetric line $f_L=f_R$, where the eigenstates of the total Hamiltonian,
$H_{\rm coup}$,  are represented as dashed lines.
The two-qubit coherent oscillations can be performed at the co-resonance point,
$f_L=f_R=0.5$. The coupled qubits are taken into the co-resonance point
from the initial position far away from it and stay for a finite time during which
temporal evolution occurs.\cite{Pashkin}

\begin{figure}[b]
\vspace*{6.6cm}
\includegraphics{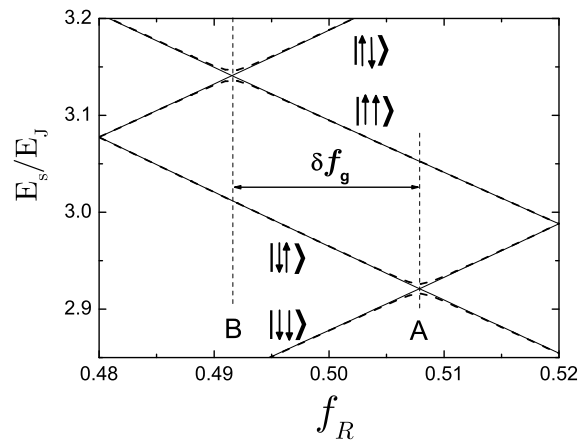}
\vspace*{-0.6cm}
\caption{Energy diagram near the resonance lines A and B when $f_L=0.48$ and
$E'_J=0.1E_J$. Solid lines represent the energies of the coupled
qubits without tunneling and dashed lines are for $t_L=t_R$=1GHz.
A and B show the values of $f_R$ where the point ($f_L$,$f_R$) is
on the resonance lines A or B.}
\label{fig:fig4}
\end{figure}

\section{Controlled-NOT operations}

We here consider the CNOT operations using  evolutions
at resonance lines\cite{Yamamoto} in Fig. \ref{fig:fig2} and show that
the large value of $J$ in present coupling scheme is able to give accurate CNOT operations.
In Fig. \ref{fig:fig4} the energy levels around lines A and B are shown.
Initially the coupled qubits are located far away from the resonance lines, say $f_R=f_L=0.457$,
being represented by the linear combination of the states,
$|\downarrow\downarrow\rangle, |\uparrow\downarrow\rangle, |\downarrow\uparrow\rangle$
and $|\uparrow\uparrow\rangle$. When we adjust  $f_R$ to the resonance line B
in Fig. \ref{fig:fig4}, the evolution
from $|\uparrow\downarrow\rangle$ to $|\uparrow\uparrow\rangle$ takes place, and vice versa.
However the states, $|\downarrow\downarrow\rangle$ and $|\downarrow\uparrow\rangle$,
do not respond to this operation.
This discriminating  operation enables one to implement the CNOT gate using the left qubit
as a control qubit and the right qubit as a target qubit.
The resonance lines C and D in Fig. \ref{fig:fig2} are also available, if we want
to use the right qubit as a control qubit.

\begin{figure}[t]
\vspace*{7cm}
\includegraphics{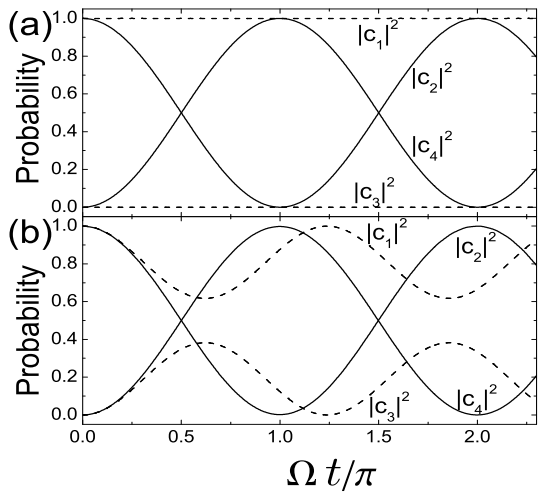}
\vspace*{-0.5cm}
\caption{Time evolutions of the coupled qubits with $t_R=t_L$=0.8GHz for
(a) $E'_J=E_J, f_R=0.457$ and $f_L=0.4837$ and
(b) $E'_J=0.005E_J, f_R=0.4995$ and $f_L=0.4793$.
The solid (dashed) lines display the probabilities that
the target (control) qubits occupy the corresponding states.}
\label{fig:fig5}
\end{figure}

\begin{figure}[b]
\vspace*{6cm}
\includegraphics{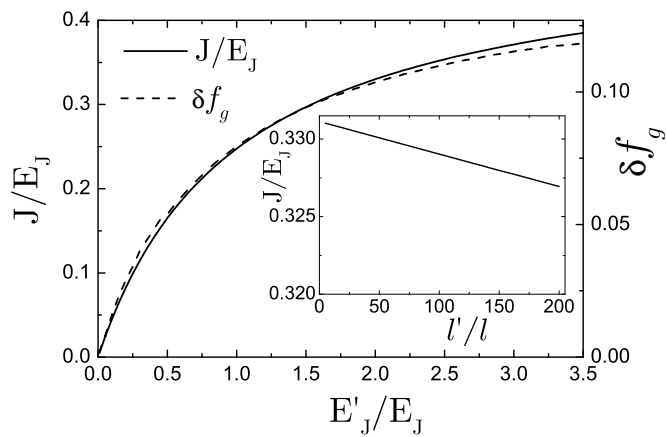}
\vspace*{-0.5cm}
\caption{ Plots of the coupling strength $J$ and the external flux difference
$\delta f_g$ versus the Josephson coupling energy in the connecting loop.
Inset shows the the coupling strength $J$
versus interqubit length $l'/l$ for $E'_J=2E_J$.
Here the coupling strength $J$ is calculated
at the co-resonance point $f_L=f_R=0.5$.}
\label{fig:DfgJL}
\end{figure}

In order to demonstrate the CNOT gate we study the time evolutions
of the coupled qubits. We start with the two qubit state written by
\begin{eqnarray}
|\Psi(t)\rangle=c_1|\downarrow\downarrow\rangle+c_2|\uparrow\downarrow\rangle
+c_3|\downarrow\uparrow\rangle+c_4|\uparrow\uparrow\rangle.
\end{eqnarray}
Considering the CNOT gate operation at resonance line B in Fig. \ref{fig:fig4},
we set the initial condition, $c_1=c_2=1$ and $c_3=c_4=0$, at $t=0$.
The time evolution of this state is obtained by the Schr{\"o}dinger equation,
$H\Psi(t)=i\hbar\partial\Psi(t)/\partial t$, with the Hamiltonian in Eq. (\ref{totalH}).
In Fig. \ref{fig:fig5} (a) we show the time evolutions of $|c_i|^2$ for the case $E'_J=E_J$
and $t_L=t_R$=0.004$E_J$=0.8GHz.
At time $\Omega t\approx\pi$ with $\Omega\equiv 2t_R=2t_L$ we can see that
the state $|\uparrow\downarrow\rangle$ evolves into the state $|\uparrow\uparrow\rangle$,
while the states, $|\downarrow\downarrow\rangle$ and $|\downarrow\uparrow\rangle$
do not respond this operation. In order to invoke the spin flip between the states
$|\downarrow\downarrow\rangle$ and $|\downarrow\uparrow\rangle$ remaining the states
$|\uparrow\downarrow\rangle$ and $|\uparrow\uparrow\rangle$ unchanged, the resonance line A
can be used for the similar process. Therefore the present scheme completes the CNOT gate operation.

For this discriminating operation the resonance lines A and B should be kept far away
from each other. When the target qubit evolves in the CNOT operation, the control qubit
must stay in the initial state. However, if the resonance lines are too close,
the unwanted oscillation of the control qubit may take place.
Thus the design parameters of the coupling scheme should be set for the value of $\delta f_g$
to be large enough to  exhibit high performance of the CNOT gate.
In  Fig. \ref{fig:fig4}, when  $f_R$ is adjusted to line B,
the amplitude of the unwanted oscillation between two states,
$|\downarrow\downarrow\rangle$ and $|\downarrow\uparrow\rangle$, can be suppressed by
the large value of $\delta f_g$ (or $J$).\cite{Yamamoto}
We obtain the values of $\delta f_g$ and $J$ as a function of
$E'_J$ as shown in Fig. \ref{fig:DfgJL}.
We also consider the small $\delta f_g$ case when $E'_J=0.005E_J$=1GHz
resulting $J$=0.49GHz to investigate how the small coupling
strength deteriorates the CNOT gate performance.
In  Fig. \ref{fig:fig5} (b) we show the time evolution of the this coupled qubits
with the same initial condition as Fig. \ref{fig:fig5} (a) and $t_L=t_R$=0.8GHz,
where we can see that the control qubit also evolves and does not stay in the initial state
when  the target qubit flips at $\Omega t\approx \pi$.
Thus we see that the CNOT gate cannot be accomplished  in this parameter regime.

\begin{figure}[b]
\vspace*{6.5cm}
\includegraphics{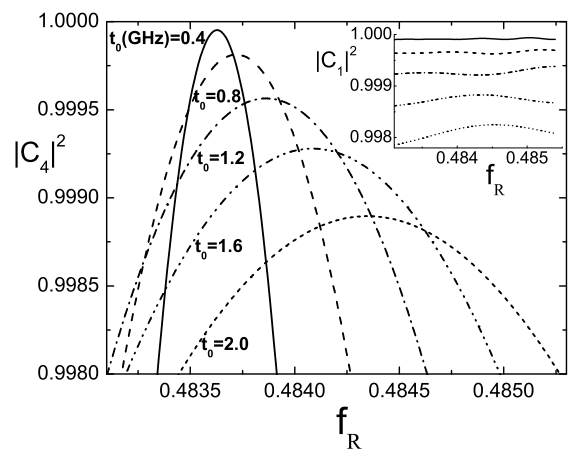}
\vspace*{-0.5cm}
\caption{Switching probabilities of the state $|\uparrow\downarrow\rangle$ to
the state $|\uparrow\uparrow\rangle$ of the coupled qubit system when $E'_J=E_J$ and $f_L=0.457$
for various tunneling amplitudes $t_L=t_R=t_0$. Inset shows the probabilities that the state $|\downarrow\downarrow\rangle$
does not respond to this operation and remains unchanged.}
\label{fig:Flip}
\end{figure}

Present coupling scheme can achieve large coupling strength by using the
phases, not flux, of the phase qubit. With the phase differences, $\phi_{L1}$ and
$\phi_{R1}$, of the Josephson junctions in Fig. \ref{fig:fig1} (a),
the effective flux has the value that
$f_{\rm eff}\equiv(\phi_{R1}-\phi_{L1})/2\pi \approx 0.32$ for $f_L\approx 0.5$
and $f_R\approx 0.5$. The value of coupling strength $J$ can be about O$(E_J)$
with $E_J$=200GHz by varying the value of $E'_J$ as shown in Fig. \ref{fig:DfgJL},
while it is just about 0.5GHz in the experiment using the mutual inductance\cite{Majer}
since the induced flux of three-Josephson junction loop is quite small\cite{Kim}
such that $f_{\rm ind}\equiv\Phi_{\rm ind}/\Phi_0 \approx 0.002$.

The remaining problem is about the accuracy of CNOT operations. By deliberately choosing the
parameter regimes for the implementation of CNOT gate, we can satisfy
the fault-tolerant criteria of quantum computation
\cite{Nielsen} which  is about $10^{-4}$. We calculate
the switching probabilities for the case in  Fig. \ref{fig:fig5} (a)
when $\Omega t\approx \pi$. In Fig. \ref{fig:Flip} the switching
probabilities for various tunneling amplitudes are shown as a function of
the external fluxes, $f_R$, with the fixed value of $f_L=0.457$ and $E'_J=E_J$.
We see in this figure that  the fidelity is higher for lower tunneling amplitude.
For higher tunneling amplitude, the evolution at a resonance line can be
affected by the presence of the other nearby resonance lines.
Actually the external fluxes in Fig. \ref{fig:Flip} correspond to the point in the sector of resonance
line B between the lines C and D where the evolution of the left qubit occurs
(Fig. \ref{fig:fig2}).  The large tunneling amplitude $t_L$ may
induce the unwanted oscillation of the controll qubit and
reduce the fidelity for the state of the target  qubit.  The slight difference of pick
points in Fig. \ref{fig:Flip}  is also attributed to this aspect.
We note that, although the probabilities are not so high as in the
low $t_0$ cases, the high $t_0$ cases have the advantage that  the
operating ranges of $f_R$ are rather broad than the low $t_0$ cases.
In the inset of Fig. \ref{fig:Flip} we show the probabilities that
the the state $|\downarrow\downarrow\rangle$ stays in the initial
state.

Many two qubit operations need the controllable coupling strength in implementation.
However, the  recent experiment on  coupled charge qubits\cite{Pashkin,Yamamoto}
and on the coupled current qubits\cite{Majer}  cannot control the coupling strength.
For  charge qubits a controllable coupling scheme using variable
electric transformer has been proposed theoretically.\cite{Averin}
Present  coupling scheme allows the tunable coupling strength for  superconducting current qubit.
The Josephson junction in connecting loop is  a dc-SQUID which is composed of two Josephson
junctions with coupling energy $E_{Jx}$. Total  Josephson coupling energy
$E'_J=2E_{Jx}\cos(\pi f_x)$ is controllable by the finite  flux $f_x$ piercing into the  dc-SQUID.
We show the coupling strength of the coupled qubits $J$ in Eq. (\ref{J}) as a function of $E'_J$
in  Fig. \ref{fig:DfgJL}.

In addition, we study the dependence of the coupling strength
$J$ on the length of the connecting loop, since
the length of each connecting loop $l'$ varies with selective coupling
in the scalable design of Fig. \ref{fig:fig1} (b).
We plot $J$  as a function of $l'/l$ in the inset of Fig. \ref{fig:DfgJL},
where $J$ remains almost constant as the value of $l'/l$ increases.
This behavior show that present scheme is appropriate for large
scale integration which requires accurate two qubit operations even for
very remote qubits.

\section{Summary}

We have proposed in this work a scheme of qubit coupling for the persistent current qubits
using a connecting loop with Josephson junctions.
The Hamiltonian of the coupled qubit system is obtained by calculating
the energy levels with the effective potential energy and the kinetic energy
originated from the charging energy of the capacitance of the Josephson junctions in qubits
and introducing the tunneling through the barrier of the double well potential.
The time resolved dynamics shows that the accurate CNOT gate operations can be
performed due to the large coupling strength which can be as large as O$(E_J)$.
Moreover the coupling strength is controllable and has a weak dependence on the length
of connecting loop of two coupled qubits selected from the scalable design.

\begin{center}
{\bf ACKNOWLEDGMENTS}
\end{center}

This work was supported by Korea Research Foundation Grant
No. KRF-2003-070-C00020.

\end{document}